# Dispersion Pre-compensation in Asymmetric Measurement Device Independent Quantum Key Distribution for Improved Secret Key Generation

**Ishan Pandey*[#], Gokul A*, Varun Raghunathan**
*Department of Electrical Communication Engineering, Indian Institute of Science, Bengaluru, India*
*U.R Rao Satellite Centre, Indian Space Research Organisation[#]*
*ishanp@iisc.ac.in, gokula@iisc.ac.in, varunr@iisc.ac.in,  *Equal Contribution*

**Abstract:** In Measurement-Device-Independent Quantum Key Distribution (MDI-QKD), key rates are significantly reduced due to dispersion in asymmetric channels. This work addresses this issue by utilizing intensity and phase modulators for dispersion compensation, thus avoiding the additional losses associated with dispersion-compensating fibers (DCFs). This approach enhances key rates and enables longer communication distances.

## 1. Introduction

In MDI-QKD systems, Alice and Bob transmit qubits to an untrusted central node, Charlie, to generate keys, which secures the system against detector side-channel attacks [1]. Alice and Bob send encoded pulses to Charlie, who performs Hong-Ou-Mandel (HOM) interference and announces the detector clicks. However, due to the asymmetric channels between Alice-to-Charlie and Bob-to-Charlie, even identical pulses from Alice and Bob experience varying degrees of dispersion, which reduces the visibility in HOM interference [2]. High HOM interference visibility is critical for distilling secret keys with low quantum bit error rates (QBER), which in turn improves the secret key rate (SKR). Achieving high visibility is challenging due to pulse width mismatches caused by asymmetric channel dispersion. Conventionally, dispersion is mitigated using dispersion-compensating fibers (DCFs); however, these introduce additional losses, ultimately reducing the achievable distance for secure key distribution. This paper presents dispersion compensation techniques using intensity and phase modulators, eliminating the need for DCFs and thereby enhancing the MDI-QKD system's robustness and extending communication range. Section 2 discusses the relationship between HOM visibility and secret key rate for MDI-QKD, Section 3 outlines the compensation methods, and Section 4 presents the implementation and results of dispersion compensation in an MDI-QKD system.

## 2. Key rate of MDI-QKD

The secret key rate for the MDI-QKD system given as follows [1]:

$$S = Qz_{11}(1 - h(e^x_{11})) - fQz_{\mu\sigma}h(ez_{\mu\sigma}) - \log_2(8/\varepsilon_{cor}) - 2\log_2(2/\varepsilon_{sec}) \qquad (1)$$

Here, $Qz_{11}$ is the single photon gain in Z basis, $e^x_{11}$ is the single photon error rate in X Basis. The term *f* represents the error correction efficiency, typically valued at 1.14. $Qz_{\mu\sigma}$ is the total gain in Z basis while $\varepsilon_{cor}$ and $\varepsilon_{sec}$ are the correction and secrecy parameters respectively.

The visibility of the channel directly influences the X-basis QBER, which consequently impacts the secret key rate (SKR) of the MDI-QKD system. This effect is illustrated in [1] by analyzing the probability of detecting a singlet state ($|\psi^-\rangle$) when both Alice and Bob send single photons from coherent sources. Visibility impacts the probability of detecting the singlet state and therefore affects the X-basis QBER.

$$P(|\psi^-\rangle|2\ Photons, visibility\ V, in) = V\ P(|\psi^-\rangle|2\ Photons, interfering, in) + (1 - V)P(|\psi^-\rangle|2\ Photons, non-interfering, in) \qquad (2)$$

The term on the left side of Equation (2) represents the total probability of singlet state detections, assuming that two photons are input to the beam splitter with visibility V across the channel. The first term on the right side describes the probability of singlet state detection under maximum interference, indicating that the two photons are perfectly identical. The last term accounts for the probability of singlet state detection when the photons have some distinguishable degree of freedom, thereby not fully interfering.

This analysis verifies that key rates are compromised due to pulse dispersion across asymmetric channels. Our compensation module significantly improves the data rate, extending the maximum secure key generation distance.

## 3. Dispersion compensation with Intensity and Phase modulator

This section explores two scenarios for implementing dispersion compensation: (i) *Known dispersion parameters*, where parameters such as the second-order dispersion coefficient and fiber lengths from Alice and Bob to Charlie are known, and (ii) *Unknown dispersion parameters*, where these parameters and fiber lengths are not known.

It is noteworthy that while the first-order dispersion coefficient shifts the dip in the coincidence detection curve, it does not degrade the curve itself. Therefore, first-order dispersion is excluded from the calculations in this paper.

### 3.1 Known Dispersion Parameters

Alice and Bob encode their qubits in time bins using Gaussian pulses of width $T_0$ as follows:

$$f_1(t) = \frac{0.7511}{\sqrt{T_0}} e^{-\frac{t^2}{2T_0^2}} \tag{3}$$

When the Gaussian pulse $f_1(t)$ undergoes dispersion, its width increases. To counteract this, a compensated Gaussian pulse $f_1^c(t)$ is transmitted, designed to match the original $f_1(t)$ after dispersion. For an optical fiber of length $z$ with a second-order dispersion coefficient $\beta^{(2)}$, the compensated pulse $f_1^c(t)$ is given by (derivation in the Appendix):

$$f_1^c(t) = 0.7511\sqrt{T_0} \frac{e^{-\frac{t^2 T_0^2}{2(T_0^4+(\beta^{(2)}z)^2)}} e^{\frac{it^2\beta^{(2)}z}{2(T_0^4+(\beta^{(2)}z)^2)} - i0.5\,tan^{-1}\left[\frac{\beta^{(2)}z}{T_0^2}\right]}}{\sqrt[4]{T_0^4+(\beta^{(2)}z)^2}} \tag{4}$$

Figure 1 shows that the shape of $f_1^c(t)$ after dispersion closely resembles $f_1(t)$, with their phases effectively swapped. The phase of $f_1(t)$, before dispersion becomes the phase of $f_1^c(t)$, and vice versa. This phase-reversed waveform enables the reconstruction of the original waveform at the receiver end when $f_1^c(t)$ is transmitted.

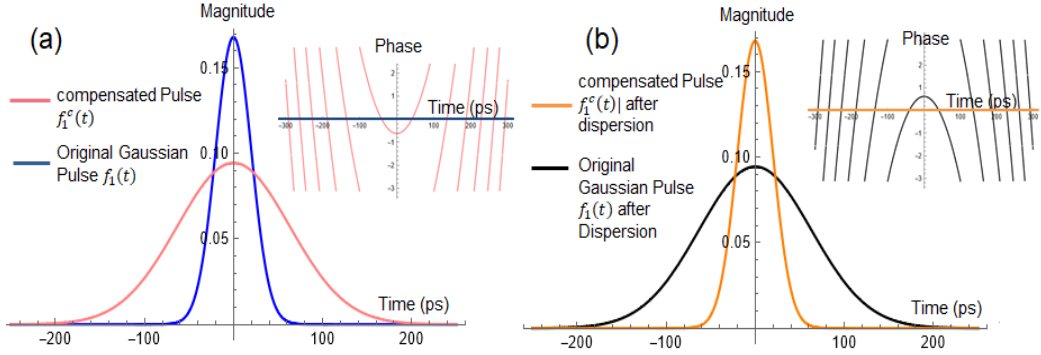

Fig. 1. Magnitude and phase plot of original gaussian pulse and compensated gaussian pulse before and after dispersion, with one arm encountering a dispersion coefficient of $\beta^{(2)} = 20 ps^2/km$ and pulse width $T_0 = 20 ps$ for (a) Input to the 60 km fiber link and (b) Output of 60 km fiber spool.

Intensity and phase modulators generate $f_1^c(t)$: the intensity modulator generates $|f_1^c(t)|^2$, while the phase modulator generates $\angle f_1^c(t)$. With known second-order dispersion coefficients $\beta_A^{(2)}$ and $\beta_B^{(2)}$ and fiber lengths $L_A$ and $L_B$, Alice and Bob transmit compensated waveforms using the values $\beta_A^{(2)} L_A$ and $\beta_B^{(2)} L_B$ in the expression for $f_1^c(t)$. As a result, the waveforms at Charlie's end, before interference, become identical, restoring the visibility and Full Width at Half Maximum (FWHM) to their expected values.

### 3.2 Unknown Dispersion Parameters

In the second scenario, where the dispersion parameters and fiber lengths are unknown, Charlie, the central node, must be trusted to honestly announce the detection events of the similar reference signals transmitted by Alice and Bob as required condition for MDI-QKD. As a result, coincidence detection curve is plotted. Ref [3] discusses the visibility of coincidence detection in the Hong-Ou-Mandel (HOM) experiment decreases due to asymmetric dispersion of the coherent state. This reduction in visibility, along with changes in the FWHM, serves as an indicator of asymmetric dispersion in the optical paths.

To estimate dispersion, Alice and Bob first send identical reference signals. Charlie, as the central node, announces the coincidence detection events, which serve as a reference for Alice and Bob to determine the visibility. With a phase-randomized source, the expected visibility in the absence of dispersion is 50%. When dispersion is present, the visibility $V$ is given by:

$$V = \frac{T_0^2}{\sqrt{4T_0^4 + \alpha^2}} \quad (5)$$

where $T_0$ represents the pulse width of the Gaussian pulse, and $\alpha = \beta_A^{(2)} L_A - \beta_B^{(2)} L_B$. Here, $\beta_A^{(2)}$ and $\beta_B^{(2)}$ are the second-order dispersion coefficients for the paths from Alice and Bob to Charlie, with fiber lengths $L_A$ and $L_B$ respectively. The FWHM value, denoted as $d$, of the HOM coincidence detection curve changes due to dispersion, providing another indicator of $\alpha$ [3]:

$$d = \sqrt{\frac{\frac{2\alpha^2}{T_0^2} + 8T_0^2}{\ln(2)}} \quad (6)$$

Table 1: Comparison of visibility, FWHM and the proposed parameter ($\Gamma$) of the coincidence detection due to HOM interference for four scenarios: without dispersion, dispersion without compensation, dispersion with compensation at Alice's end and dispersion with compensation at Bob's end for unknown dispersion parameters.

|  | Visibility | FWHM | $\Gamma$ |
|---|---|---|---|
| *Without Asymmetric dispersion* | 0.5 | 67.9457 | 0 |
| *Asymmetric Dispersion without compensation* | 0.3989 | 148.87 | 8.1813 |
| *Asymmetric Dispersion and compensation at Bob's end* | 0.4901 | 72.248 | 0.0424 |
| *Asymmetric Dispersion and compensation at Alice's end* | 0.4615 | 119.236 | 1.9759 |

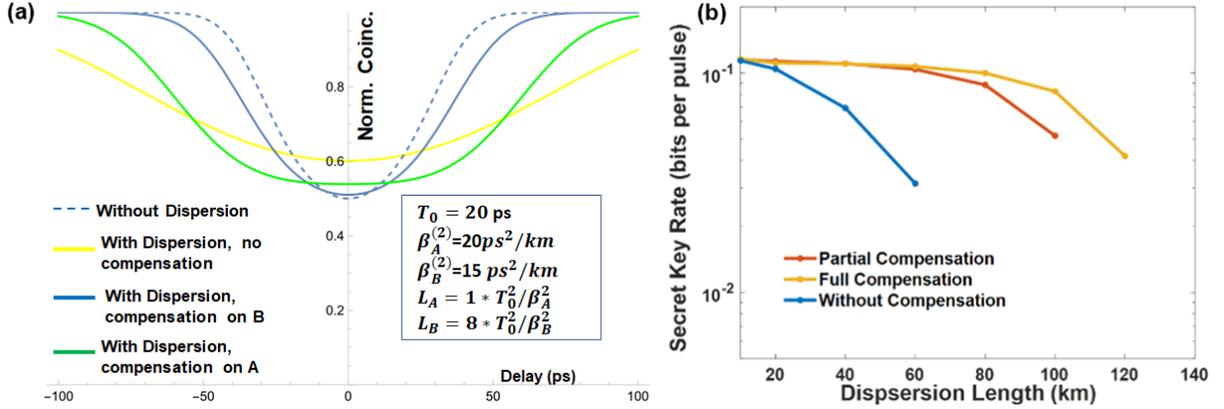

Fig. 2. (a) Coincidence detection curve for asymmetric dispersion at fiber from Alice and Bob to Charlie. (b) Secret Key Rates vs Dispersion length with and without dispersion compensation.

To identify which path encounters greater dispersion, Alice and Bob each attempt to compensate for their reference signals individually by substituting the value of $\alpha$ in place of $\beta^{(2)} z$ in the expression for $f_1^c(t)$. Thereby, two more time windows will be utilized for sending reference signal with compensated reference signal waveform transmitted by one party at a time. This process results in two coincidence curves, one with compensation of Alice and other with compensation of Bob.

For each curve, a parameter $\Gamma$ is calculated as:

$$\Gamma = |FWHM_{obtained} - FWHM_{expected}| \times |Visibility_{obtained} - Visibility_{expected}| \quad (7)$$

The expected values of FWHM and visibility are calculated by setting $\alpha=0$ in equations (5) and (6), assuming symmetric conditions or no dispersion in either path from Alice or Bob to Charlie. By comparing the $\Gamma$ values derived from two coincidence detection curves—one for each path where compensation is applied—the path with the minimum $\Gamma$ value is selected for final implementation, as shown in Table 1.

This chosen compensated setup is then utilized for secure key distribution. Figure 2(a) illustrates the coincidence detection curve, showing that, with proper compensation, the curve closely aligns with the ideal, dispersion-free curve. However, if the difference in dispersion between the two paths is minimal, the proposed method may yield unpredictable results. This limitation suggests an area for future work: refining compensation techniques to perfectly match the dispersion-free curve.

## 4. Results and Discussion

The pre-compensation module significantly improves key generation rates in MDI-QKD systems, as seen in Fig. 2(b), where a theoretical model [1] shows up to a 3.4-fold increase over a 60 km channel. This improvement arises from dispersion mitigation, which typically reduces key rates. Additionally, the compensation module extends the secure key generation link length. Compared to partial compensation, full compensation maximizes visibility. This study focuses on enhancing MDI-QKD links with intensity and phase modulators to improve HOM interference visibility, increase SKR, and extend communication range without DCFs. The setup supports QKD networks in star topologies, where Charlie serves as the central node for parties seeking shared secret keys, enhancing SKR even with unknown dispersion parameters.

## 5. Conclusion

In conclusion, this study demonstrates that intensity and phase modulators can effectively mitigate dispersion in MDI-QKD, eliminating the need for DCFs. This approach improves HOM interference visibility, enhances SKR, and extends secure key generation distances, enabling robust quantum communication over extended ranges.

## 6. References

[1] Lo H K, Curty M and Qi B 2012 Measurement-device-independent quantum key distribution *Phys. Rev. Lett.* **108** 130503.

[2] S. K. Ranu, A. Prabhakar, and P. Mandayam, "Effect of Pulse-shape Mismatch on the Security of Measurement-device-independent QKD Protocols," in *Frontiers in Optics + Laser Science 2022 (FIO, LS)*, Technical Digest Series (Optica Publishing Group, 2022), paper JW4B.29.

[3] Y. Fan, C. Yuan, R. Zhang, S. Shen, P. Wu, H. Wang, H. Li, G. Deng, H. Song, L. You, Z. Wang, Y. Wang, G. Guo, and Q. Zhou, "Effect of dispersion on indistinguishability between single-photon wave-packets," Photon. Res. 9, 1134-1143 (2021).

[4] Govind P. Agrawal, Non Linear Fiber Optics, Academic Press.

[5] Zhiyu Chen, Lianshan Yan, Hengyun Jiang, Wei Pan, Bin Luo, and Xihua Zou, "Dispersion Compensation in Analog Photonic Link Utilizing a Phase Modulator" JOURNAL OF LIGHTWAVE TECHNOLOGY, VOL. 32, NO. 23, DECEMBER 1, 2014.

## APPENDIX 1: Derivation of pre-compensated waveform

To find the expression of pre-compensated pulse ($f_1^c(t)$) such that after it encounters dispersion, the resulting pulse becomes the same as what we are willing to transmit. The pulse width of a Gaussian pulse ($f_1(t) = \frac{0.7511}{\sqrt{T_0}} e^{-\frac{t^2}{2T_0^2}}$) pre-compensated by a variable x such that pre-compensated pulse becomes:

$$f_1^c(t) = A \cdot f_1(t - x) \quad A1$$

Here, A is a constant to ensure the energy of $f_1(t)$ is same as $f_1^c(t)$. Therefore, to find the value $A$ from equating their energies, $\int_{-\infty}^{\infty} |f_1(t)|^2 dt = \int_{-\infty}^{\infty} |f_1^c(t)|^2 dt$, we get

$$A = \frac{T_0}{T_0 - x} \quad A2$$

The $f_1^c(t)$ undergoes second order dispersion to become a pulse defined by $f_2^c(t)$ such that:

$$f_2^c(t) = IFT\left[FT[f_1^c(t)] \cdot e^{i\beta^{(2)} \cdot z \cdot \frac{\omega^2}{2}}\right] = \frac{0.7511\sqrt{T_0} e^{\frac{t^2}{2i\beta^{(2)}z - 2(T_0 - x)^2}}}{\sqrt{(T_0 - x)^2 - i\beta^{(2)}z}} \quad A3$$

Here, $\beta^{(2)}$ is the second order dispersion coefficient, $FT$ and $IFT$ are Fourier transform and inverse Fourier transform respectively. As we want that pulse after passing through dispersive media becomes same as what we are willing to receive. Hence, equating $f_2^c(t)$ with $f_1(t)$ to find the value of $x$ as:

$$x = T_0 \pm \sqrt{T_0^2 + i\beta^{(2)}z} \quad A4$$

Taking the value of $x = T_0 - \sqrt{T_0^2 + i\beta^{(2)}z}$, we get $f_1^c(t) = 0.7511\sqrt{T_0} \frac{e^{-\frac{t^2 T_0^2}{2(T_0^4 + (\beta^{(2)}z)^2)}} e^{\frac{it^2 \beta^{(2)} z}{2(T_0^4 + (\beta^{(2)}z)^2)} - i0.5 \tan^{-1}\left[\frac{\beta^{(2)}z}{T_0^2}\right]}}{\sqrt[4]{T_0^4 + (\beta^{(2)}z)^2}}$ .